\documentclass[prx,aps,longbibliography,showpacs,groupedaddress,superscriptaddress,twocolumn,toc=flat,nofootinbib]{revtex4-1}

\usepackage{graphicx}
\usepackage{latexsym}
\usepackage{amssymb}
\usepackage{amsmath}
\usepackage{amsfonts}
\usepackage{enumerate}
\usepackage{upgreek}
\usepackage{subfigure}
\usepackage{bm}
\usepackage{nicefrac}
\usepackage{bbold}
\usepackage{verbatim}
\usepackage[unicode=true,
 bookmarks=false,
 breaklinks=false,pdfborder={0 0 1},backref=false,colorlinks=true]
 {hyperref}
\hypersetup{
 linkcolor=[rgb]{0,0,1},citecolor=[rgb]{0,0,1},urlcolor=[rgb]{0,0,1}}
\usepackage{multirow}
\usepackage{color}
\usepackage{comment}
\usepackage{xcolor}
\usepackage{soul}
\usepackage{tikz}
\usepackage{braket}

\newcommand{\C}{\hat{c}}
\newcommand{\Cd}{\hat{c}^\dagger}

\newcommand{\n}{\hat{n}}
\newcommand{\Ham}{\hat{\mathcal{H}}}
\newcommand{\hc}{\mathrm{H.c.}}

\newcommand{\nn}{\langle \mathbf{i}, \mathbf{j}\rangle}
\renewcommand{\i}{\mathbf{i}}
\renewcommand{\j}{\mathbf{j}}
\renewcommand{\u}{\uparrow}
\renewcommand{\d}{\downarrow}
\newcommand{\Dint}{\mathcal{D}}
\newcommand{\PF}{\mathcal{Z}}
\newcommand{\bPhi}{\mathbf{\Phi}}
\newcommand{\bO}{\mathbf{\Omega}}
\renewcommand{\bell}{\boldsymbol{\ell}}
\renewcommand{\k}{\mathbf{k}}
\newcommand{\zz}{\mathbb{Z}}

\newcommand{\bbrakket}[1]{\langle\!\langle #1 \rangle\!\rangle}

\AtBeginDocument{%
    \newwrite\bibnotes
    \def\bibnotesext{Notes.bib}
    \immediate\openout\bibnotes=\jobname\bibnotesext
    \immediate\write\bibnotes{@CONTROL{REVTEX41Control}}
    \immediate\write\bibnotes{@CONTROL{%
    apsrev41Control,author="08",editor="1",pages="1",title="0",year="1"}}
     \if@filesw
     \immediate\write\@auxout{\string\citation{apsrev41Control}}%
    \fi
}%

\begin{document}
\title{Geometric orthogonal metals: Hidden antiferromagnetism \\ and pseudogap from fluctuating stripes}

\author{Henning Schl\"omer}
\email{h.schloemer@physik.uni-muenchen.de}
\affiliation{Department of Physics and Arnold Sommerfeld Center for Theoretical Physics (ASC), Ludwig-Maximilians-Universit\"at M\"unchen, Theresienstr. 37, M\"unchen D-80333, Germany}
\affiliation{Munich Center for Quantum Science and Technology (MCQST), Schellingstr. 4, M\"unchen D-80799, Germany}

\author{Annabelle Bohrdt}
\affiliation{University of Regensburg, Universitätsstr. 31, Regensburg D-93053, Germany}
\affiliation{Munich Center for Quantum Science and Technology (MCQST), Schellingstr. 4, M\"unchen D-80799, Germany}

\author{Fabian Grusdt}
\email{fabian.grusdt@physik.uni-muenchen.de}
\affiliation{Department of Physics and Arnold Sommerfeld Center for Theoretical Physics (ASC), Ludwig-Maximilians-Universit\"at M\"unchen, Theresienstr. 37, M\"unchen D-80333, Germany}
\affiliation{Munich Center for Quantum Science and Technology (MCQST), Schellingstr. 4, M\"unchen D-80799, Germany}

\date{\today}

\begin{abstract}
One of the key features of hole-doped cuprates is the presence of an extended pseudogap phase, whose microscopic origin has been the subject of intense investigation since its discovery and is believed to be crucial for understanding high-temperature superconductivity. Various explanations have been proposed for the pseudogap, including links to symmetry-breaking orders such as stripes or pairing, and the emergence of novel fractionalized Fermi liquid (FL*) and orthogonal metal (OM) phases. The topological nature of FL* and OM phases has been identified as scenarios compatible with a small Fermi surface without symmetry breaking, as suggested experimentally. With recent experimental and numerical studies supporting an intricate relationship between stripe order and the pseudogap phase, we here propose an alternative scenario: an orthogonal metal with a geometric origin (GOM) driven by fluctuating domain walls. The essential mechanism behind our proposal is hidden order, where the proliferation of domain walls stabilized by charge fluctuations obscures the underlying long-range antiferromagnetic order in real-space, but order is preserved in the reference frame of the background spins. As a result, well-defined fermionic quasiparticles in the form of magnetic polarons exist, which couple to $\zz_2$ topological excitations of the domain wall string-net condensate in the ground state and constitute a small Fermi surface. At a critical doping value, we argue that hidden order is lost, driving a transition to a regular Fermi liquid at a hidden quantum critical point (hQCP) featuring quantum critical transport properties. Our GOM framework establishes a deep connection between the antiferromagnetic, stripe, and pseudogap phases, and suggests a possible unification of superconductivity in (electron and hole) doped cuprates and heavy fermion compounds.       
\end{abstract}
\maketitle

\section{Introduction}
The discovery of high-temperature superconductivity in doped cuprates~\cite{Bednorz1986} has shaped the field of contemporary condensed matter physics for almost four decades. Tremendous progress has been made in understanding their rich phase diagram, including the exotic normal phases from which superconductivity arises; nonetheless, a unified understanding remains elusive~\cite{Lee2006, Keimer2015}. 

One particular conceptual puzzle of the cuprates' phase diagram concerns the microscopic origin of the pseudogap phase~\cite{Timusk1999, Norman2005, Chowdhury2015}. It is characterized by a partial depletion of low-energy excitations, most prominent in spectroscopic observables, resulting in notorious ``Fermi arcs''~\cite{Shen2005, Kanigel2006, Yang2011, Kurokawa2023}. While photoemission experiments suggest the absence of coherent fermionic quasiparticles, quantum oscillations advocate the opposite, i.e., the existence of Fermi-liquid-like fermionic quasiparticles moving on closed semiclassical orbits around small Fermi pockets~\cite{Doiron2007, Vignolle2008, Yelland2008, Ramshaw2011}. This is corroborated by Hall~\cite{Ando2004, Badoux2016, Collington2017}, optical conductivity~\cite{Sayed2013} and magnetoresistence~\cite{Chan2014} measurements, showing vanilla Fermi liquid behavior consistent with a small Fermi surface, i.e., with carrier density $\delta$ (where $\delta$ is the hole doping away from the Mott insulating state), and not $1+\delta$ as expected from Luttinger's theorem~\cite{Luttinger1960}. The latter relates the volume enclosed by a Fermi surface to the density of microscopic charge carriers. 

While antiferromagnetic (AFM) order that breaks the lattice translational symmetry constitutes a simple possible explanation for several characteristics of the pseudogap phase~\cite{Chowdhury2015, Sachdev_QPM}, it is inconsistent with experimental data across broad families of cuprates, where only short-range AFM correlations that span a few lattice sites are observed beyond $\sim 5\%$ doping~\cite{Lee2006, Xu2009}. Nevertheless, spin-wave like excitations (paramagnons) with dispersions and spectroscopic characteristics similar to those of magnons in antiferromagnetic (AFM) Mott insulators exist in the entire pseudogap phase~\cite{LeTacon2011, Dean2013, Guarise2014}. The question of how doping concentrations of a few percent can diminish long-range AFM order in such an efficient way while preserving spin-wave-like excitations remains another unsolved problem. 

Here, we propose that the mechanism behind the destruction of AFM order upon doping is \textit{hidden order}, associated with sublattice fluctuations. While the SU(2) symmetry of the background spins is spontaneously broken, fluctuations of charges linked to AFM domain walls result in the order being obscured in the lab frame. Moreover, we present a scenario where quantum fluctuations of such domain walls lead to the existence of topological excitations in the ground state, resulting in the formation of a small (fractionalized) Fermi surface constituted by magnetic (or spin-) polarons. Before describing our approach in more detail, we briefly review the current theoretical understanding of the pseudogap phase. 

In low-temperature regions of the phase diagram, adjacent to the pseudogap regime, symmetry-breaking orders such as the stripe phase appear, defined by static spin- and charge order~\cite{Emery1999, Kivelson2003, Vojta2009}. They are particularly prominent in Lanthanum-based compounds, with pronounced signals around $\delta=1/8$ doping~\cite{Tranquada1995, Tranquada1996, Abbamonte2005}. More subtle stripe-like signatures, predominantly in the charge sector, have further been shown to exist in Bismuth~\cite{Howald2003, Vershinin2004, Parker2010} and Yttrium-based~\cite{Wu2011} cuprate families through scanning-tunneling-microscope (STM) and nuclear magnetic resonance (NMR) measurements. In particular, by applying magnetic fields that suppress superconductivity, stripe order can be induced in samples where static order without additional fields is absent~\cite{Wu2011, Chang2009}. 

The observation of stripes in broad classes of copper oxides has motivated extensive studies of the relation between the pseudogap phase and symmetry breaking order~\cite{Vojta2012}. This is supported by NMR measurements on various La-based compounds, which indicate that intertwined spin- and charge stripes are inherently present up to a critical doping $\delta_c$, beyond which stripe order as well as signatures of the pseudogap phase vanish~\cite{Frachet2020, Vinogard2022, Missiaen2024}. This finding is further reinforced by recent numerical simulations of the Fermi-Hubbard (FH) model, which suggest that the doping and interaction regimes where pseudogap and stripe phases emerge at high and low temperatures, respectively, align precisely~\cite{Simkovic2024, Xu2022_stripes}.

On the one hand, it has been put forward that the nature of quantum oscillations in the pseudogap at low temperature is linked to translational symmetry breaking, whereby density-wave order reconstructs the (large) conventional Fermi surface into small electron pockets~\cite{Millis2007, Dimov2008, Norman2010, Yao2011}. In a similar spirit, the existence of a quantum critical point as a function of doping has been proposed to explain the anomalous transport properties found in the high-temperature region of the pseudogap regime, i.e. linear-in temperature resistivity~\cite{Castellani1995, Kivelson1998}. Here, the corresponding critical doping corresponds to the point where the critical temperature of a putative symmetry-breaking order that competes with superconductivity vanishes. It has been argued, however, that the disruption of a large Fermi surface by conventional (thermally fluctuating) order unlikely explains universal Fermi liquid-like observations with carrier density $\delta$ in the pseudogap phase~\cite{Chowdhury2015, Punk2015}. One particular objection is that the transition into the pseudogap at temperature $T^*$ does not seem to be captured by a thermodynamic phase transition, though Berezinskii–Kosterlitz–Thouless-type transitions with weak thermodynamic signatures can not be ruled out~\cite{Vojta2012}.

On the other hand, the pseudogap has been described as a precursor of superconductivity, whereby the formation of incoherent, preformed Cooper pairs leads to a partial depletion of fermionic spectral weight. While support of this idea has been reported in a variety of experiments~\cite{Wang2006, Yang2008, Kanigel2008, niu2024e}, temperatures where evidence of preformed pairs has been found are significantly below the pseudogap temperature $T^*$. Furthermore, recent measurements of iridates revealed that the pseudogap can exist in broad parameter regimes without the appearance of superconductivity at lower temperatures, which supports a disparate nature of the the two phases~\cite{Battisti2017, Hsu2024}.   

This has led to a third class of proposed scenarios, where the symmetry-breaking orders found at low temperatures are interpreted as instabilities of a distinct ``pseudogap'' phase of matter~\cite{Chowdhury2014, Chowdhury2015, Punk2015}. A prominent scenario is the formation of a fractionalized Fermi liquid (FL$^*$)~\cite{Senthil2003}, which features well-defined fermionic quasiparticles in the absence of symmetry-breaking order while violating Luttinger's theorem. In the single-band Hubbard model, this scenario can be realized when assuming that local moments form an odd spin liquid with topological excitations (visons)~\cite{Read1991, Wen1991}, into which mobile hole carriers are doped~\cite{Kaul2007, Qi2010, Moon2011, Mei2012, Punk2012tJ, sachdev2016}. While dopants in a spin liquid can decay into fractionalized spin and charge constituents~\cite{Kotliar1988}, assuming bound states of the latter leads to a metal with a small Fermi surface~\cite{Punk2015}. On technical grounds, this apparent violation of Luttinger's theorem is caused by topological excitations of the spin liquid, which can absorb momentum in Oshikawa's flux insertion protocol~\cite{Oshikawa2000} that corresponds exactly to unit density (leading to a Fermi surface of volume $\propto \delta$ instead of $\propto 1+\delta$).

An intimately related phase is the orthogonal metal (OM)~\cite{Ruegg2010, Nandkishore2012, Gazit2020}, where, in contrast to the FL$^*$, the fermionic quasiparticles are gauge-charged and therefore orthogonal to the physical electron. In this scenario, transport and quantum oscillation measurements continue to reflect the presence of a small Fermi surface. However, in the ground state, this surface does not appear in ARPES due to the absence of coherent electron quasiparticles. Nevertheless, as we discuss later, incoherent, broadened features in the electronic spectral function can still emerge at finite temperature---consistent with experimental observations in cuprates.

\begin{figure}
\includegraphics[width=0.9\columnwidth]{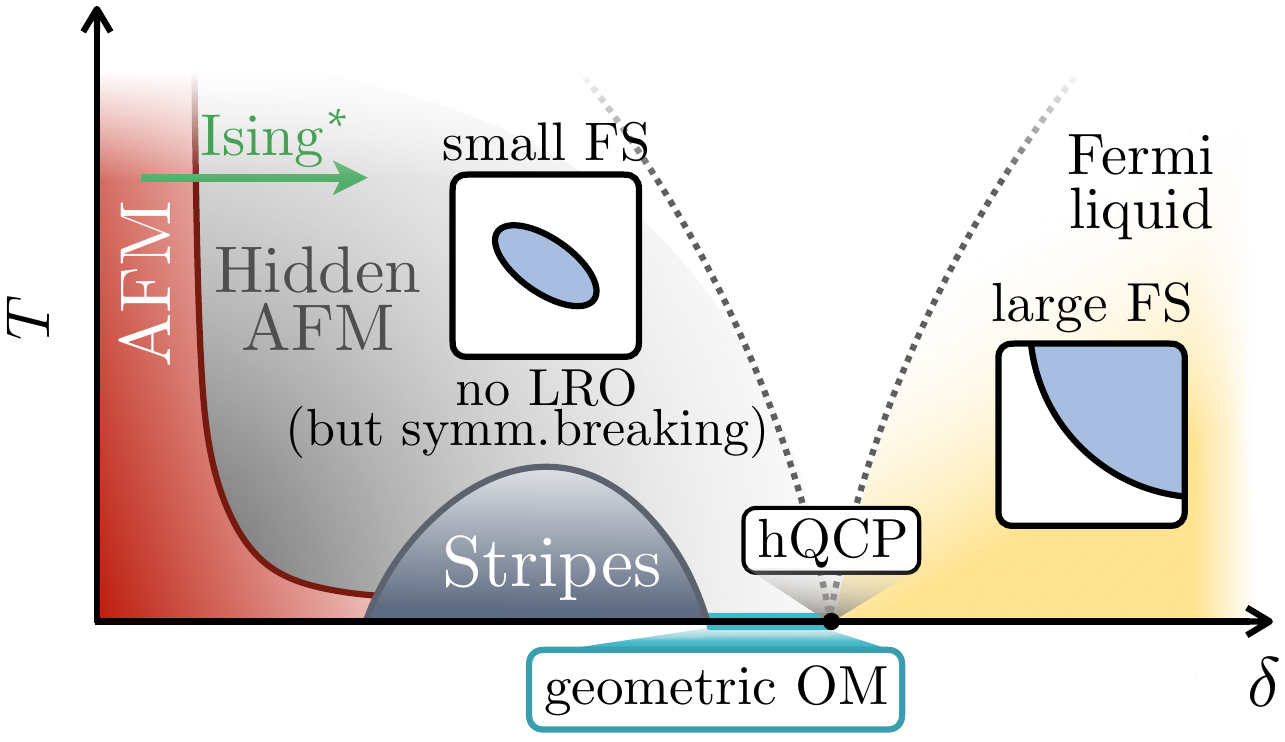}
\caption{\textbf{Phase diagram of hole-doped cuprates.} Interpretation of the phase diagram of hole-doped cuprates at strong magnetic fields, where superconductivity is suppressed. Around half-filling, long-range AFM order exists. Upon doping, the string tension $h$ of domain wall loops decreases, driving an Ising* transition to a hidden ordered state (green arrow). At zero temperature, we propose that cuprates realize a geometric orthogonal metal (GOM), whereby topological excitations of the fluctuating domain walls lead to the formation of a small Fermi surface constituted by magnetic (or spin-) polarons. At a critical doping, hidden order vanishes through a frustration mechanism governed by fluctuations of microscopic holes, resulting in a hidden quantum critical point (hQCP) that drives a topological transition from a GOM (with a small Fermi surface) to a Fermi liquid (with a large Fermi surface).}
\label{fig:fig1}
\end{figure}

\subsection{Overview of our results} Our description of the pseudogap phase builds upon the idea that domain walls of the underlying AFM -- i.e. individual stripes -- can be held together by strong spin-charge correlations, even in a regime where interactions between neighboring stripes can be overcome by thermal or quantum fluctuations. This prevents the formation of charge-density wave order and results in a low-energy theory of string-like fluctuating stripes that leads us to propose a microscopic theory for an orthogonal metal phase: A string-net condensate of stripes introduces topological order which lends an emergent gauge charge to the magnetic polaron excitations of the underlying AFM. This fractionalizes their Fermi surface and geometrically suppresses long-range AFM correlations, in a state with an underlying spontaneously broken SU(2) symmetry. In particular, our proposed scenario consists of the following, see Fig.~\ref{fig:fig1}:
\begin{enumerate}[(i)]
    \item The mechanism behind the efficient disruption of the AFM phase close to half-filling is \textit{hidden order}, caused by fluctuating domain walls of the spin-background linked to charge fluctuations. Specifically we propose that, while the SU(2) symmetry of the AFM background is broken, the order is hidden through fluctuations of closed loops that enclose areas where the AFM order switches its sublattice. The transition between visible and hidden order is governed by an Ising* criticality characterized by highly non-local observables.
    \item Hidden order allows for the existence of well-defined fermionic quasiparticles, i.e., magnetic (or spin-) polarons, as well as bosonic Goldstone modes (paramagnons). The latter live in the space of the ordered spin background, referred to as squeezed space.
    \item At low temperatures, interactions between fluctuating domain walls drive a transition to the stripe phase, where individual stripes align and excitations are gapped out. In this case, magnetic order reappears in the lab frame through long-range incommensurate spin correlations.
    \item At zero temperature and enhanced doping (and strong magnetic fields), where neither stripes nor superconductivity appear in the ground state, the pseudogap phase is the realization of a loop gas of fluctuating domain walls. String-net condensation leads to hidden order and the formation of an odd $\zz_2$ spin liquid. 
    \item Magnetic polarons couple to the geometric spin liquid formed by the fluctuating domain walls, acquire an emergent gauge charge, and form a small Fermi surface. The broken SU(2) symmetry together with sublattice fluctuations define what we call a geometric orthogonal metal (GOM), which features topologial excitations that absorb momentum in Oshikawa's flux insertion protocol.
    \item At a hidden quantum critical point (hQCP) on the doping axis, hidden order disappears and the SU(2) symmetry is fully restored as spin-interactions in squeezed space become frustrated. Signatures of this transition are detectable only through highly non-local observables or quantum critical transport properties. We conjecture these to underlie the bad metallicity observed in cuprates. 
\end{enumerate}

The picture of hidden order allows to naturally unite the AFM and stripe phases with the pseudogap: In all cases, the SU(2) symmetry of the spin background is spontaneously broken; it is merely hidden in the experimentalist's frame of reference within the pseudogap phase. 
We argue that by postulating hidden order in the system, a geometric fractionalized orthogonal metal constituted by magnetic polarons naturally emerges. It further directly suggests the presence of spin-wave-like excitations (paramagnons)~\cite{LeTacon2011, Xu2009} which, in contrast, should be absent when the physical spins in the system form the spin liquid~\cite{Punk2015}. 

We emphasize the distinction between our proposed scenario for the pseudogap and the ``fluctuating stripes'' framework introduced earlier (see, e.g., Refs.~\cite{Kivelson2003, Zaanen2001}). In our approach, we propose that fluctuating antiferromagnetic (AFM) domain walls form a non-local string-net condensate, giving rise to emergent topological excitations. Crucially, our theory does not involve competing orders, nor does it associate the pseudogap with quantum critical fluctuations. This distinction is further highlighted by the nature of the domain walls: whereas the ``fluctuating stripes'' scenario is limited to unidirectional, fluctuating stripes, our framework envisions domain walls that ``bend'' and form an isotropically fluctuating string-net structure.

Due to the inherent real-space nature of relevant observables, hidden order can be directly probed with ultracold atom experiments. Groundbreaking innovations in the field of ultracold atoms in optical lattices~\cite{Bloch2008, Esslinger2010, Bloch2012, Parsons2015, Cheuk2015, Haller2015, Gross2017} have led to a plethora of insights into the intermediate-temperature regime of the doped Fermi-Hubbard (FH) model in recent years~\cite{Bohrdt2020}, including the observation of magnetic order~\cite{Greif2013, Hart2015, Boll2016, Mazurenko_AFM, shao2024} as well as the formation of magnetic polarons~\cite{Koepsell_nature2019, Chiu2019, Hartke2020, Koepsell2021, Ji2021} and stripe-like structures~\cite{bourgund2023}. By looking for signs of hidden order in spin-resolved snapshots of analog quantum simulation platforms, our scenario can be explored with existing experimental setups. This further stresses the potential of analog simulation of the FH model at finite temperatures, which may help to differentiate between various theoretical proposals of the pseudogap phase and pin down its microscopic nature. 

\section{The geometric OM scenario}

In the following we motivate the origin of hidden antiferromagnetism and the geometric orthogonal metal starting from the microscopic two-dimensional (2D) FH model, which is broadly believed to capture the essential low-energy physics of cuprates~\cite{Anderson1987}. In its particle-hole symmetric formulation, the Hamiltonian reads \\ \\
\begin{equation}
\begin{aligned}
    \Ham_{\text{FH}} = -t \sum_{\nn} &\sum_{\sigma=\u, \d} \Cd_{\i, \sigma} \C_{\j, \sigma} + \hc \\ &+ U \sum_{\i} \left(\n_{\i, \u}-\frac{1}{2}\right) \left(\n_{\i, \d}-\frac{1}{2}\right),
\end{aligned}
\end{equation}
where $\hat{c}_{\mathbf{i}, \sigma}^{(\dagger)}$ and $\hat{n}_{\mathbf{i}}$ are fermionic annihilation (creation) and particle density operators on site $\mathbf{i}$, respectively; $\braket{\mathbf{i}, \mathbf{j}}$ denotes nearest neighbor (NN) sites on the 2D square lattice. Anticipating a metallic state close to an AFM instability at half filling, the system can be decoupled in the particle-hole channel by introducing the bosonic collective mode $\bPhi$ (Hubbard-Stratonovich field), often referred to as the ``paramagnon field''~\cite{Sachdev_QPM}. The (exact) partition function then reads

\begin{widetext}
\begin{equation}
\begin{aligned}
    \PF = \int \prod_{\i, \sigma} \Dint c_{\i \sigma}(\tau) \prod_{\i} \Dint \bPhi_{\i}(\tau) \exp \Bigg( -\int d\tau \Bigg\{ \sum_{\k, \sigma} \bar{c}_{\k, \sigma} \Big[ \frac{\partial}{\partial \tau} + \epsilon_{\k} \Big] c_{\k, \sigma} + \sum_{\i} \left[ \frac{3}{8U} \bPhi_{\i}^2 - \bPhi_{\i} \cdot \mathbf{S}_{\i} \right] \Bigg\}  \Bigg),
    \label{eq:pf}
\end{aligned}
\end{equation}
\end{widetext}
where we have used the local identity $U \left(\n_{\i, \u}-\frac{1}{2}\right) \left(\n_{\i, \d}-\frac{1}{2}\right) = -\frac{2U}{3} \mathbf{S}_{\i}^2 + \frac{U}{4}$ with the spin defined as $\mathbf{S}_{\i} = \frac{1}{2} \sum_{\alpha,\beta} \bar{c}_{\i, \alpha} \boldsymbol{\sigma}_{\alpha, \beta} c_{\i ,\beta}$. Though acquiring the full solution of Eq.~\eqref{eq:pf} is an extremely challenging task, a systematic treatment of the path integral can yield insights into the phases of the FH model in certain limits. 

Condensation of paramagnons leads to a spin-density wave instability with wave vector $\mathbf{K} = [\pi, \pi]$ on the square lattice, resulting in a metallic state with spontaneous spin polarization that has opposite orientation $\bPhi_{\i \in A}/|\bPhi_{\i \in A}| = - \bPhi_{\i \in B}/|\bPhi_{\i \in B}|$ on $A$ and $B$ sublattices, i.e., $\bPhi_{\i} = \bO_{\i} \exp(i \mathbf{K} \cdot \mathbf{r}_{\i})$ with $\braket{\bO_{\i}}\neq 0$ and $\bO_{\i}$ the alignment of the N\'eel field. A long-wavelength field theory (the ``spin-fermion'' model) then describes Fermi surface reconstruction due to translational symmetry breaking. In particular, hole pockets are formed close to hot-spots where the spin-density wave gaps out the electronic spectrum~\cite{Chowdhury2015, Sachdev_QPM}. This realizes the well-understood scenario where the pseudogap emerges from a spontaneously broken translational symmetry~\cite{Eberlein2016}. In this case, a strong-coupling perspective of the pseudogap is provided by a theory of magnetic (or spin-) polarons, which form hole pockets around $\mathbf{q} = [\pi/2, \pi/2]$, consistent with low-energy charge excitations at low doping~\cite{Shraiman1988, Sachdev1989, Kane1989, Emery1987, Schrieffer1988, Beran1996, Grusdt2018_partons, Bohrdt2020_parton, Bermes2024}.

Previously, it has been argued that a fractionalized Fermi liquid (FL*) can be realized by restricting fluctuations of $\bPhi$ only to its angle~\cite{Qi2010, Moon2011}. The spin-density wave can then become ``quantum disordered''~\cite{Chowdhury2015}, stabilizing an exotic state of small hole pockets but without AFM order, i.e.~$\braket{\bO_{\i}} = 0$. Nevertheless, the local magnitude of AFM order is finite, which can be formalized by a Higgs field in an SU(2) gauge theory after transformation of the underlying electrons to a rotating frame~\cite{Sachdev2009, Chowdhury2015, sachdev2016, Sachdev2016_sdw, Chatterjee2017, Sachdev2019_review, Bonetti2022}. Similarly, an FL* phase can be realized when the background spins form a quantum spin liquid e.g. through resonances of spin-singlet dimers (analytically represented by emergent gauge fields)~\cite{Chowdhury2015}. This is in particular motivated by exploring possible ground states of the undoped parent Hamiltonian (e.g. a $\zz_2$ spin liquid), which is then doped to form a fractionalized Fermi liquid under certain assumptions~\cite{Punk2015}.

In the following, we take an alternative approach and include stripe-like AFM domain wall defects in an effective description of the FH model. These correspond to spatially well-localized sign-changes ($\pi$-phase slips, $\bO \rightarrow -\bO$) across line-like defects corresponding to individual (fluctuating) stripes~\cite{Zhang2002competing}, see Fig.~\ref{fig:fig2}~(a). 
Fluctuations of these defects will then lead us to the geometric orthogonal metal (GOM) scenario. Before discussing the latter, we describe how fluctuating domain wall structures efficiently hide AFM correlations in real-space, which is the essential mechanism behind the GOM at low temperatures and explains the quick demise of AFM at temperatures above the stripe ordering transition. 

\subsection{Hidden order from fluctuating stripes} 
\begin{figure}
\includegraphics[width=\columnwidth]{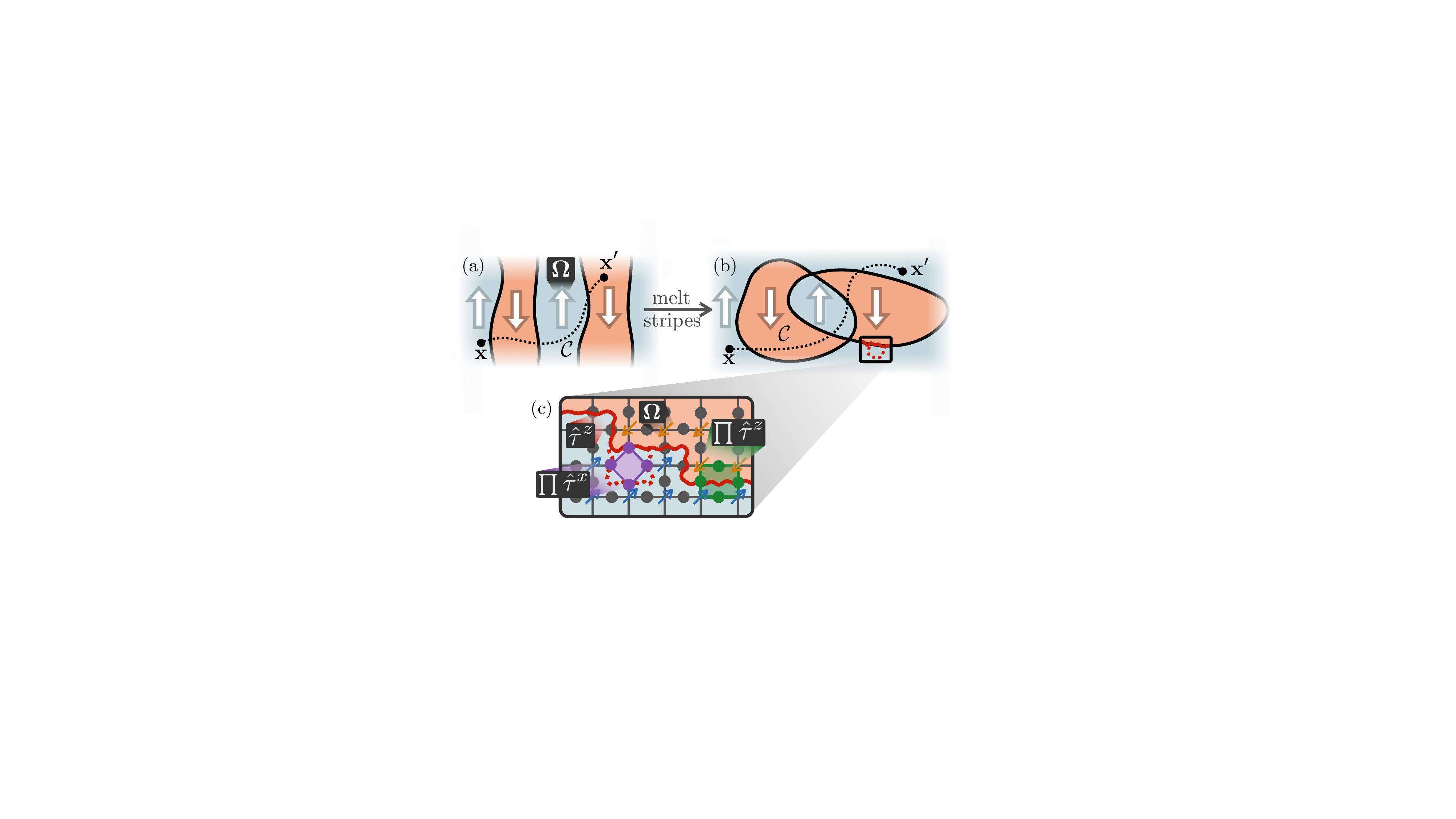}
\caption{\textbf{Fluctuating stripes.} (a) Schematic illustration of ordered stripes, breaking the discrete C$_4$ lattice symmetry. Along lines of enhanced hole density, $\pi$-phase slips of the AFM order parameter $\bO$ appear, illustrated by the arrows. (b) Upon melting stripes, we propose a scenario where domain walls form closed loops and fluctuate. While magnetic order is obscured in real space, hidden order survives, which can be revealed through non-local spin-string-spin correlations by explicitly considering the domain walls that occur along a path $\mathcal{C}$ [dashed lines in (a) and (b)], cf. Eq.~\eqref{eq:HO}. (c) We describe fluctuating domain walls by introducing strings (red wiggly line) $\hat{\tau}^z$ that live on the links (grey dots) of the lattice. The classical Hamiltonian Eq.~\eqref{eq:H_cl} describes thermal fluctuations of domain wall loops and includes plaquette terms $\propto \prod \hat{\tau}^z$ of the $\hat{\tau}^z$-field (green square). Adding quantum fluctuations $\propto \prod \hat{\tau}^x$ (purple diamond) to the Hamiltonian leads to an effective description of fluctuating domain walls through a perturbed toric code, Eq.~\eqref{eq:TC}.}
\label{fig:fig2}
\end{figure}

We consider low-energy contributions to Eq.~\eqref{eq:pf} that are directly motivated by the low-temperature phases of the FH model at finite doping: In the strongly interacting limit $U\gg t$, all state-of-the-art numerical methods broadly agree on the appearance of stripes in the ground state, where charges accumulate around AFM domain walls and form a charge density wave~\cite{Machida1989, Kato1990, Zaanen_stripes, George_stripes, White_stripes, White_stripes2, Zheng2017, Huang2018, Jiang_hubbard_pd, Qin_absence_SC, Xu2024}. This results in long-range charge and incommensurate long-range magnetic order, in accordance with broad experimental evidence in cuprate materials~\cite{Tranquada1995, Tranquada1996, Abbamonte2005}. Though quantum fluctuations of the microscopic holes lead to deformations of the lines of stripes, they are locked in place, i.e., individual stripes align and explicitly break the C$_4$ symmetry of the underlying square lattice. This scenario is schematically illustrated in Fig.~\ref{fig:fig2}~(a), showing a representative low-energy configuration in the stripe phase. Here, the indicated arrows represent the direction of $\bO$, i.e., they specify the sublattice parity~\cite{Zaanen2001} of the underlying AFM state.

Upon raising the temperature, stripes melt, and the FH model displays a pseudogap with only short-range AFM correlations present~\cite{preuss1997, Macridin2006, Kyung2006, Ferrero2009, Sordi2012, Wu2018, Simkovic2024} and a reconstructed Fermi surface~\cite{Osborne2020}. We particularly highlight Ref.~\cite{Simkovic2024}, where the fate of the pseudogap in the FH model when tuning the temperature towards the ground state has been analyzed through a `handshake' of wave-function and finite temperature based methods. Quoting Ref.~\cite{Simkovic2024}: {\it The range of density and coupling strength where a pseudogap is found precisely coincides with that in which ground-state studies find a stripe phase with long-range spin and charge order.} This further aligns with experimental findings of various La-based cuprates with different balances between superconductivity and stripe order, where \textit{the overarching conclusion is that the pseudogap and stripe phases are closely linked}~\cite{Missiaen2024}. This suggests an intricate interplay between the stripe and pseudogap phase in the FH model, with a possible common microscopic origin -- it is this scenario that we work out in depth in the remainder of this paper. 

Specifically, we now propose a set of microscopic configurations that may be relevant for an effective description of the doped FH model at elevated temperatures. To this end, we note that the formation of stripes can be associated with two energy scales: (i) The energy gain for binding a doped hole into an (undirected) $(d-1)$-dimensional AFM domain wall structure, i.e., sub-dimensional regions of accumulated charge density across which the AFM develops $\pi$-phase slips, and (ii) the interaction strength between such individual stripes, resulting in the formation of charge-density waves where individual stripes lock into place in the ground state. Generally, it is conceivable that either of the two energy scales is larger, which then governs the important contributions to Eq.~\eqref{eq:pf} at intermediate temperatures. 

We shall here focus on the scenario where the first energy scale is larger, i.e., breaking up stripe-like domain wall structures is associated with a higher energy cost compared to them locking into place. This is generally expected at low doping where the typical distance between individual stripes is large, and is supported by latest cold-atom experiments which suggest the formation of fluctuating, stripe-like structures at elevated temperatures, on the order of $J$, in (mixed-dimensional) FH systems~\cite{bourgund2023}.

We specifically focus on closed loops that enclose regions where the AFM order parameter is flipped, which we argue constitute a minimal scenario to describe the low-energy physics of domain wall fluctuations at temperatures above the stripe ordering transition. Our scenario of closed domain wall loops can further be motivated by energetic considerations: opening a closed domain wall string--thereby creating two string endpoints--incurs a magnetic energy cost on the order of $J$, as it forces two neighboring spins into alignment. Further separating these endpoints is expected to increase the energy monotonically with their distance, providing an efficient mechanism that favors the persistence of closed loops. An exemplary configuration is shown in Fig.~\ref{fig:fig2}~(b); any closed loop encloses a region where the AFM order parameter $\bO$ is sign-flipped. 

We note that in the ground state, it has been established that kinetic effects lead to directed stripes being energetically favorable over undirected configurations~\cite{Eskes1998, Zaanen_stripes}. However, at finite temperatures a scenario where the charge-density wave stiffness vanishes and loop defects proliferate is conceivable, as proposed in Ref.~\cite{Kruger2002}. From a microscopic point of view, we propose that mobile holes in the form of magnetic polarons form a Fermi sea, and local spin-charge correlations arising from microscopic interactions stabilize closed-loop domain wall configurations of the AFM background.

Our starting point is hence a state with broken SU(2) symmetry, where $\bO$ points in a particular direction; thermal fluctuations of the AFM background are described through a non-linear $\sigma$ model (NLSM)~\cite{Haldane1983}. On top of this we allow for closed loop domain wall excitations where the AFM order parameter is flipped, i.e., $\bO \rightarrow -\bO$.
Our aim is to simplify the full description of the field $\bO(\mathbf{x})$ to string configurations $\hat{\tau}(\mathbf{x})$ that describe the $\pi$-phase slips of $\bO$. To this end, we introduce operators $\hat{\tau}^z_{\bell}$ that live on the links of the lattice, where $\bell = \braket{\i,\j}$ is the link that connects sites $\i$ and $\j$. For sharp domain walls of width $\lesssim a$ (with $a$ the lattice constant), we define 
\begin{equation}
    \hat{\tau}^z_{\bell = \braket{\i, \j}} = \text{sgn}(\bO_{\i} \cdot \bO_{\j}).
    \label{eq:tauz}
\end{equation}
These local $\zz_2$ degrees of freedom then capture the structure of domain walls on the two-dimensional square lattice, illustrated in Fig.~\ref{fig:fig2}~(c).  

Focusing on classical (thermal) fluctuations for now, this motivates the following Hamiltonian to describe the thermodynamic properties of closed line domain walls at elevated temperatures, 
\begin{equation}
\hat{\mathcal{H}}_{\text{cl}} =  - K_{\square} \sum_{\square}\prod_{\bell \in \square} \hat{\tau}^z_{\bell} - h \sum_{\bell} \hat{\tau}^z_{\bell}, 
\label{eq:H_cl}
\end{equation}
where the sum runs over all plaquettes on the square lattice and $\hat{\tau}_{\bell}^z \ket{q}_{\bell} = \pm \ket{q}_{\bell}$ with $q = 0,1$ a $\zz_2$ degree of freedom defined on links $\bell$. The $K_{\square}>0$ term energetically favors an even number of strings per plaquette, which suppresses open string configurations and favors closed loop structures [cf. the green plaquette in Fig.~\ref{fig:fig2}~(c)]; $h$ takes the role of a chemical potential, i.e. it relates to the density of strings, and corresponds to a linear string tension of the loops (with energy $2h|\Sigma|$ where $|\Sigma|$ is the length of the loop). Assuming that hole dopants provide the glue that stabilize AFM domain walls, i.e. 
\begin{equation}
1-\braket{\hat{\tau}^z_{\bell}} \propto \delta,  
\label{eq:tau_delta}
\end{equation}
we can expect $h$ to decrease with increasing doping $\delta$. 

We now analyze thermal fluctuations of the classical model in Eq.~\eqref{eq:H_cl}; the effect of quantum fluctuations and emergence of a small Fermi surface is addressed further below. In the subspace of closed loops (i.e. in the pure gauge scenario $K_{\square} \rightarrow \infty$), the Hamiltonian Eq.~\eqref{eq:H_cl} is dual to the 2D Ising model~\cite{Peierls1936, wegner1971duality}, 
\begin{equation}
    \hat{\mathcal{H}}_{\text{Ising}} = -h \sum_{\braket{\i,\j}} \hat{\sigma}^z_{\i} \hat{\sigma}^z_{\j} + \text{const.},
\end{equation}
where $\hat{\sigma}^z_{\i}$ are qubit operators that live on sites $\i$ of the lattice. By construction of the duality mapping, $\hat{\sigma}^z_{\i} = \text{sgn}(\bO_{\i} \cdot \bO_0)$, where $\bO_0$ denotes the N\'eel order parameter at an arbitrary reference site. We thus see that line-defects of $\pi$-phase slips in an AFM with spontaneously broken SU(2) symmetry lead to an effective Ising description of the N\'eel order parameter. 

\begin{figure}
\includegraphics[width=0.8\columnwidth]{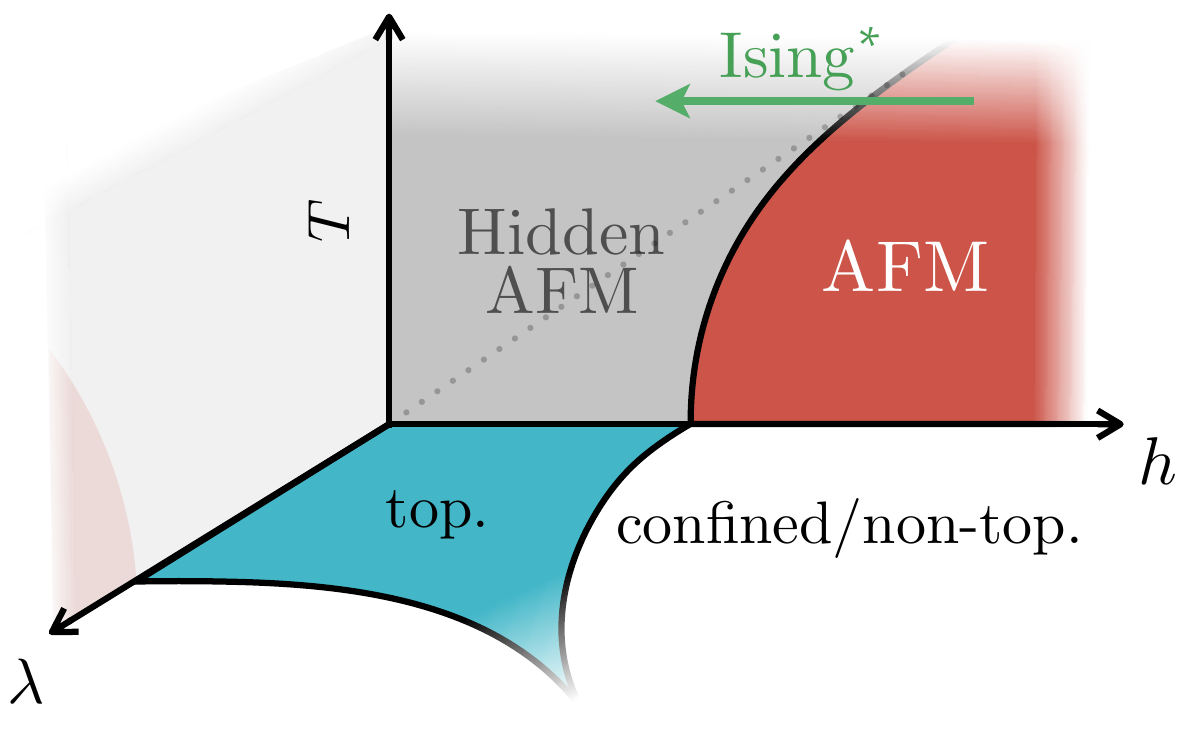}
\caption{\textbf{Phase diagram of the perturbed toric code Hamiltonian.} (a) Schematic phase diagram of Eq.~\eqref{eq:TC}. We propose a scenario in which doping leads to the formation of fluctuating closed loop domain walls across which the AFM order parameter switches its sign, leading to an extended phase that features \textit{hidden order}. At elevated temperatures, we argue that thermal fluctuations of string loops are captured by the (classical) Hamiltonian Eq.~\eqref{eq:H_cl}, which features a dual Ising criticality (Ising$^*$) that describes a transition from long-range AFM to hidden AFM order due to percolating string nets (green arrow). This corresponds to the regime in the $T$-$h$ plane where $T_c/h = \text{const}.$, illustrated by the dotted grey line. At low temperatures, quantum fluctuations stabilize the hidden order phase, characterized by loop condensation (blue area).}
\label{fig:fig3}
\end{figure}

The dual Ising Hamiltonian $\hat{\mathcal{H}}_{\text{Ising}}$ has a phase transition at the critical temperature $T_c/h \approx 2.27$, which separates a ferromagnetic from a paramagnetic phase. In terms of the Hamiltonian Eq.~\eqref{eq:H_cl}, for $T<T_c$ domain wall loops are confined and non-percolating; in the high-temperature phase, $T>T_c$, strings deconfine, i.e., loops span the whole system and percolate~\cite{Linsel2024}. The Ising ferromagnetic (paramagnetic) phase of the dual model can hence be identified with a long-range (short-range) ordered $\bO$ field in the doped Mott insulator. Though no local order parameter can be defined in terms $\hat{\tau}_{\bell}^z$ that characterizes the confined-deconfined phase transition of Eq.~\eqref{eq:H_cl}, it is governed by an Ising criticality through the dual mapping (and is hence denoted by  Ising$^*$)~\cite{wegner1971duality}.

The high-temperature deconfined phase defines the concept of \textit{hidden order} in our scenario: While real space correlations of $\bO$ become short range for $T>T_c$, the order is merely hidden efficiently through the existence of a percolating string net. In other words, for a given snapshot of the physical system, one can reconstruct the loop configuration by tracking the sublattice parity of each spin. This is akin to hidden order in 1D and mixed-dimensional systems, which has been formalized through the notion of squeezed space~\cite{Kruis2004, OgataShiba}, where the latter captures the non-local structure of correlations~\cite{Hilker2017, Grusdt2018, Grusdt2020, Schloemer2023_robust, Schloemer2023_quantifying}. 

In Ref.~\cite{Zaanen2001}, squeezed space has been generalized to uni-directed stripes in two dimensions. Sublattice parity order has been identified as the stripes' defining characteristic, which is defined by long-range properties of the non-local, topological correlation function (here written in the continuum limit and using our link variable $\hat{\tau}_{\bell}^z$ introduced above):
\begin{equation}
    \mathcal{O}_{\text{top}}(\mathbf{x} - \mathbf{x}', \mathcal{C}) = \braket{\Psi|\bO(\mathbf{x}) e^{i\pi \int_{\mathcal{C}} d\mathbf{y} [1-\hat{\tau}^z_{\braket{\mathbf{y}, \mathbf{y}+d\mathbf{y}}}]} \bO(\mathbf{x}')|\Psi},
    \label{eq:HO}
\end{equation}
where $\mathcal{C}$ is a path connecting points $\mathbf{x}$ and $\mathbf{x}'$. The contour integral ensures that domain walls are captured in the magnetic correlations, see Fig.~\ref{fig:fig2}~(a), and it becomes independent on the choice of $\mathcal{C}$ (i.e. topological in nature) when $\hat{\tau}^z$'s from closed loops. When all dopants are bound into stripes, $\hat{\tau}^z$ in Eq.~\eqref{eq:HO} can be expressed in terms of the electron density operator $\hat{n}(\mathbf{y})$, with different relations for fully and partially filled stripes~\cite{Zaanen2001}. 

The notion of hidden order through fluctuating, closed domain wall loops is a natural extension of this concept: By identifying loops of flipped AFM order parameter, one can reconstruct the original symmetry broken state with long-range order by flipping the spins within each closed loop, as illustrated in Fig.~\ref{fig:fig2}~(b). Using spin- and charge-resolved snapshots, cold-atom quantum simulators at state-of-the-art temperatures can directly search for these structures by evaluating Eq.~\eqref{eq:HO}, which highlights the potential of exploring the intermediate temperature regime of doped antiferromagnets with ultracold atoms~\cite{Bohrdt2020}.

\begin{figure*}
\includegraphics[width=\textwidth]{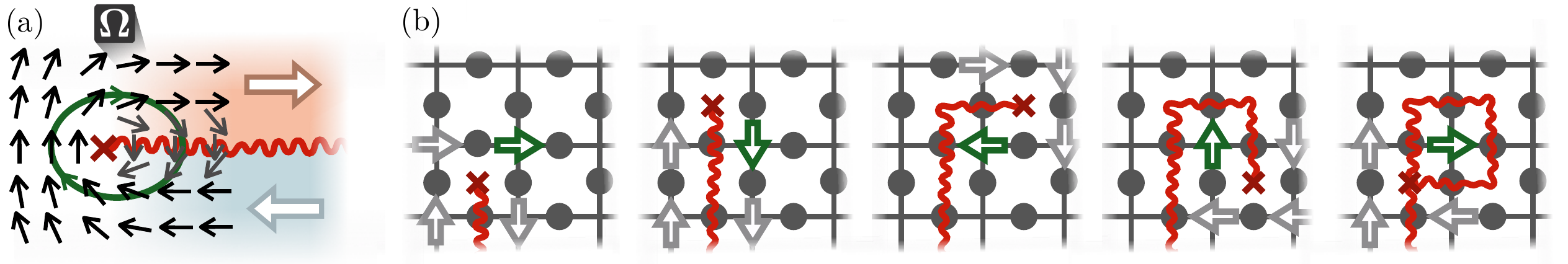}
\caption{\textbf{Berry phases.} (a) Winding of the $\bO$ field around an open domain wall string, corresponding to a half-vortex excitation. We identify these excitations as magnetic visons in the effective toric code description. Across the domain wall, the discontinuity of $\bO$ is described by a rotation in the continuum limit on scales around the lattice spacing, see the grey arrows. When adiabatically following a closed path that encircles an open string end, the field performs a full $2\pi$ rotation - leading to a Berry phase of $\pi$ when a spin-$1/2$ defect (e.g. a magnetic polaron) winds around an open domain wall end (along the green circle). (b) Effect of adiabatically moving an open string end (unit $\zz_2$ magnetic charge) around a lattice site. Encircled microscopic spins perform an in-plane rotation by $2\pi$. This leads to the lattice sites carrying a $\zz_2$ electric charge, i.e., there exists a background unit charge that ultimately leads to the formation of an odd $\zz_2$ spin liquid.}
\label{fig:fig4}
\end{figure*}

The phase diagram of the Hamiltonian Eq.~\eqref{eq:H_cl} is shown in the upper part of the $T-h$ plane in Fig.~\ref{fig:fig3}. Here, the regime where classical fluctuations [described by Eq.~\eqref{eq:H_cl}] are expected to dominate the physics (and quantum effects can be neglected) corresponds to the part of the phase diagram where $T_c/h = 2.27$, see the dotted line in Fig.~\ref{fig:fig3}.

We propose that this mechanism of hidden AFM correlations ultimately leads to the fast disappearance of the AFM phase in doped cuprates. At elevated temperatures, the Ising$^*$ criticality drives a transition from the AFM phase with long-range order to the hidden AFM phase with only short-range correlations in real space. A schematic phase diagram of doped cuprates at strong magnetic fields, i.e. where the $d$-wave superconducting phase is suppressed, is presented in Fig.~\ref{fig:fig1}. Starting at half-filling, doping leads to an increase of string density Eq.~\eqref{eq:tau_delta}, in turn reducing the string tension in the loop gas. We can therefore expect that the doping level $\delta$ of the microscopic system and the string tension $h$ of the effective classical model~Eq.~\eqref{eq:H_cl} are inversely related, resulting in a drop of the critical temperature of the AFM phase in Fig.~\ref{fig:fig1}. The transition to the hidden order phase at elevated temperatures is then governed by the same Ising$^*$ criticality as in Eq.~\eqref{eq:H_cl}, as indicated in both Fig.~\ref{fig:fig1} and Fig.~\ref{fig:fig3} by the green horizontal line. This picture is consistent with experimental signatures of an Ising-type order parameter found in the pseudogap regime~\cite{Li2008Unusual}.

Upon decreasing temperature, we propose that a transition from hidden order to long-range, incommensurate stripe order appears, whereby individual stripes lock into place and break both translational and rotational lattice symmetries. This corresponds to an ordering transition of the $\hat{\tau}^z$ fields which gaps out the spectrum, and is akin to mean-field ground state considerations in Ref.~\cite{Zaanen2001}. 

To summarize, the partition function $\mathcal{Z}$ in Eq.~\eqref{eq:pf} so far is expressed in terms of a path integral over fermions $c$ coupled to the N\'eel-order parameter $\bO$. Above we have argued that the relevant low-energy configurations in the pseudogap regime correspond to domain-wall lines of the N\'eel order, supplemented by long-wavelength fluctuations of the hidden AFM state. This motivates the expression of the partition function, after integrating out fermions, as
\begin{equation}
    \mathcal{Z} = \int \prod_{\i} \mathcal{D} \tilde{\bO}_{\i}(\tau) \prod_{\bell} \mathcal{D} \tau^z_{\bell}(\tau) e^{ S_{\rm eff} }.
    \label{eqZcl}
\end{equation}
Here $\tilde{\bO}$ is the hidden N\'eel order field in (the 2D generalization of) squeezed space, and $\tau^z$ describes the line-like AFM domain wall defects; together these determine $\bO$. So far the effective action $S_{\rm eff}$ consists of the NLSM for $\tilde{\bO}$ and the thermal fluctuations governed by the classical loop-gas Hamiltonian from Eq.~\eqref{eq:H_cl}.

In the above discussion of the loop gas, we have focused on zero matter density, i.e., we excluded open ends of domain wall strings in our description. In this setting, the percolating (hidden order) phase extends up to infinite temperature, though the stiffness and correlation length of the underlying SU(2) symmetry broken state (governed by a NLSM) is reduced with rising temperature. 
In particular at elevated temperatures (and akin to previous considerations of fluctuating stripes~\cite{Eskes1998, Zaanen2001, Kruger2002, Zhang2002competing}), we expect that including open domain wall string excitations constitutes an important step towards a more accurate description of the fluctuating stripe scenario. Ends of strings then correspond to topological half-vortices of the $\bO$ field, illustrated in Fig.~\ref{fig:fig4}~(a). We speculate that explicitly taking into account open ends and their interactions may lead to a BKT-type transition from the hidden order to a disordered phase, driven by the unbinding of (half) vortex-antivortex pairs connected by domain wall strings. Such a transition would mark the edge of the pseudogap phase observed at $T^*$, but will be discussed in more detail in a forthcoming work.

\subsection{Berry phases and emergent toric code} Although we restricted our considerations to closed domain wall loops so far, analyzing the braiding statistics of topological vortex charges will give us important insights into the nature of gauge fluctuations that need to be included in the effective low-temperature theory that we construct next. This will ultimately lead us to a $\zz_2$ lattice gauge theory (LGT) with unit background charge, in turn realizing an odd $\zz_2$ quantum spin liquid in the ground state.

As a starting point, we revisit the step from Eq.~\eqref{eq:pf} to Eq.~\eqref{eqZcl} in which we have focused on classical terms and ignored Berry phase effects. While this is justified at elevated temperatures, including Berry phase terms associated with topological defects of $\bO$ is crucial to obtain the correct low-temperature description. As above, our goal is to express the microscopic field $\bO$ in terms of the smoothly varying hidden N\'eel field $\tilde{\bO}$ -- for which Berry phases are not important at low doping -- and the fluctuating string field $\tau^z$ describing domain walls. The latter contains information about topological excitations of $\bO$ and must therefore reflect the associated quantized Berry phases.

To understand how this can be achieved, we discuss the gauge structure of $\tau^z$ and clarify how it relates to $\bO$. In our description of string configurations $\tau^z$, domain wall lines can be interpreted as $\zz_2$ magnetic field lines on the dual lattice. Ends of domain wall lines give rise to topological half-vortex excitations of the $\bO$ field, where the parity of the AFM order parameter continuously changes its sign. This is depicted in Fig.~\ref{fig:fig4}~(a) for a particular rotation direction on the Bloch sphere (we note, however, that in the path integral, the field is integrated over all rotation directions, which restores the global $S_{\text{tot}}^z$ symmetry via NLSM thermal fluctuations of the hidden antiferromagnet). 

We therefore identify the ends of domain wall lines as vison excitations (or $\zz_2$ magnetic charges) of an effective $\zz_2$ LGT theory.
We now ask what happens when adiabatically moving such a vison around a site $\i$ of the physical lattice, see Fig.~\ref{fig:fig4}~(b). Due to the vortex structure of the associated $\bO$-configuration, this winds the spin on site $\i$ once around itself, Fig.~\ref{fig:fig4}~(b). Thereby, the many-body wave function picks up a topological Berry phase of $\varphi_B = \Omega_S/2 = \pi$, as one microscopic spin-$1/2$ completes a full rotation around the Bloch sphere and hence encloses a solid angle of $\Omega_S = 2\pi$. 

Next we construct the effective action of the $\tau^z$ field. To this end we include quantum fluctuations and make sure that the ground state of the field $\tau^z$ inherits the $\pi$-Berry phase derived above, produced by the microscopic field $\bO$ that we integrate out in going from Eq.~\eqref{eq:pf} to Eq.~\eqref{eqZcl}. Microscopically, quantum fluctuations of domain wall loops originate from fluctuating mobile dopants; here we capture them by the simplest effective Hamiltonian consistent with the topology of closed loops. 

We introduce the operator $\hat{\tau}^x_{\bell}$ conjugate to the $\hat{\tau}^z_{\bell}$ that flips the $\zz_2$ degree of freedom, $\hat{\tau}^x_{\bell} \ket{1}_{\bell} = \ket{0}_{\bell}$. Including terms $\prod_{\ell\in +_{\i}} \hat{\tau}^x_{\bell}$ in the Hamiltonian, where $+_{\i}$ includes all links around a site $\i$, thus constitutes a minimal scenario: When applying flips on the purple links in Fig.~\ref{fig:fig2}~(c), the red domain wall loop is extended to surround the site connected by the purple links -- thus allowing the domain wall to fluctuate without introducing open strings. Within this scenario, fluctuating stripes are described by a perturbed toric code Hamiltonian~\cite{Kitaev2003},
\begin{equation}
\begin{aligned}
    \hat{\mathcal{H}}_{\text{TC}} = &- K_{\square} \sum_{\square}\prod_{\bell \in \square} \hat{\tau}^z_{\bell} - h \sum_{\bell} \hat{\tau}^z_{\bell} \\ &+ K_{+} \sum_{+}\prod_{\bell \in +} \hat{\tau}^x_{\bell}  -\lambda \sum_{\bell} \hat{\tau}^x_{\bell}.
\end{aligned}
\label{eq:TC}
\end{equation}
As in the case of classical fluctuations, we in this work focus on the limit of fully closed domain-wall loops. Therefore we set $\lambda = 0$ in the following, although we note that small non-vanishing $\lambda$ would not change our conclusions below. 

What remains to be clarified is the sign of $K_+$. To this end we note that the choice $K_+>0$ that we make fixes the gauge sector to $\prod_{\bell \in +}\hat{\tau}^x_{\bell} = -1$, ensuring a unit $\zz_2$ electric background charge. The mutual braiding statistics of a vison around a $\zz_2$ electric charge corresponds to a $\pi$-Berry phase in the topologically ordered ground state of the perturbed toric code Hamiltonian Eq.~\eqref{eq:TC}. Therefore we conjecture that integrating out $\bO$-configurations associated with AFM domain wall defects introduces a large, positive $K_+$ term in the effective Hamiltonian of the fluctuating domain-wall field $\tau^z$ that correspond to a topologically ordered ground state whose vison excitations carry the required $\pi$-Berry phase of half-vortex defects, see Fig.~\ref{fig:fig4}~(b).

The well-known phase diagram of Hamiltonian Eq.~\eqref{eq:TC} as a function of $\lambda, h$ and $T$ (assuming $K_{\square} = K_{+}$) is shown in Fig.~\ref{fig:fig3} -- see also Refs.~\cite{wen2004,Levin2005,Trebst2007, Wu2012}. Within the blue region at $T=0$, the ground state features topological order, realizing an odd $\zz_2$ spin liquid due to the non-vanishing background charge~\cite{Read1991}. In particular, this regime corresponds to a deconfined phase characterized by loop gas condensation~\cite{wen2004, Levin2005}, whereby quantum fluctuations lead to the extension of the hidden AFM phase to the ground state. 

While stripes are expected to be stabilized by weak string-string interactions [not explicitly included in Eq.~\eqref{eq:TC}] at zero temperature, we propose that the normal state with a small Fermi surface observed around optimal doping (and in the presence of a magnetic field to suppress superconductivity)~\cite{Badoux2016} realizes a topologically ordered hidden AFM as described above, see Fig.~\ref{fig:fig1}. 

\subsection{Geometric OM} This finally brings us to the analysis of charge carriers in the hidden AFM phase. We argue that these are constituted by magnetic (or spin-) polaron excitations~\cite{Shraiman1988, Sachdev1989, Kane1989, Emery1987, Schrieffer1988, Beran1996, Grusdt2018_partons, Bohrdt2020_parton, Bermes2024} of the hidden AFM, with a charge carrier density given by the number of free dopants $\delta$. This suggests that they form a small Fermi surface, and as we will show next by analyzing their coupling to gauge fluctuations, this is in accordance with the generalized Luttinger theorem for fractionalized Fermi liquids~\cite{Oshikawa2000, Senthil2003, Chowdhury2015} when the AFM domain walls $\tau^z$ form a topologically ordered string-net condensate. 

From this perspective, the ordered stripe phase breaks the translational symmetry, which can lead to the formation of charge- and (incommensurate) spin-density wave orders~\cite{Chowdhury2014, Tranquada1995, Tranquada1996, Abbamonte2005}. Note that even in the hidden AFM phase with fluctuating AFM domain walls we expect strong spin-charge correlations to stabilize such domain walls. However, this does not require the charge carriers to be bound into these fluctuating structures, see also Ref.~\cite{Xu2024}; our picture is rather that they simultaneously participate in a small-FS Fermi sea while stabilizing the loop gas of the domain wall strings. 

Our effective field theory in Eq.~\eqref{eqZcl} is still lacking the degrees of freedom associated with mobile dopants. At very low doping, before domain-wall strings proliferate, the AFM breaks the SU(2) and translational symmetries of the system, and its elementary charged excitations correspond to magnetic polarons. In the following we will denote them by fermionic operators $\hat{f}_{\i,\tilde{\sigma}}$ carrying spin-$1/2$ and charge $e$. The index $\tilde{\sigma}=\pm 1$ labels the two N\'eel-ordered vacuum states related by time-reversal; e.g. $\tilde{\sigma}=+1$ corresponds to $\uparrow \downarrow \uparrow \downarrow \uparrow \dots$, and $\tilde{\sigma} = -1$ to $\downarrow \uparrow \downarrow \uparrow \downarrow \dots$. Together with the sub-lattice of $\mathbf{i}$, this determines the spin $\sigma$ carried by the magnetic polaron.

We now include magnetic polarons in the more general partition function in Eq.~\eqref{eq:pf} by introducing an additional path integration $\int \prod_{\i} \mathcal{D} f_{\i,\tilde{\sigma}}(\tau)$. In the parent AFM, corresponding to $\hat{\tau}^z \equiv 1$ everywhere, individual hole dopants lead to the formation of magnetic polarons, which gain kinetic energy through spin-exchange processes in the underlying quantum magnet. This leads to an effective next-to-nearest neighbor (NNN) hopping of magnetic polarons and a corresponding dispersion that has its minimum around $[\pi/2,\pi/2]$~\cite{Shraiman1988, Sachdev1989, Kane1989, Emery1987, Schrieffer1988, Beran1996, Grusdt2018_partons, Bohrdt2020_parton, Bermes2024}. The microscopic spin $\sigma = \pm 1$ (corresponding to the spin-up and spin-down states) of the magnetic polaron in a background with fluctuating AFM domain walls is determined by the sub-lattice index $\alpha=\pm 1$ (for A- and B-sublattice, respectively), the vacuum label $\tilde{\sigma}$ as well as the string configuration $\{\hat{\tau}^z\}$, which we label by $\Sigma$, 
\begin{equation}
\sigma(\alpha, \tilde{\sigma}, \Sigma) = \alpha \tilde{\sigma} (-1)^{\Sigma}.
\label{eq:sigsigp}
\end{equation}
The latter introduces a sign denoted $(-1)^{\Sigma}$ corresponding to being ``inside'' or ``outside'' of closed loops formed by links $\bell$ with $\hat{\tau}^z = -1$, i.e., it describes the parity of the loop gas at a given spatial position. Note that in the parent AFM, $(-1)^{\Sigma} \equiv +1$ everywhere. 

Moving through a non-uniform AFM, the spin of a magnetic polaron adiabatically follows the surrounding N\'eel field $\bO(\mathbf{x})$. This has important consequences when AFM domain walls, described by the field $\tau^z$, proliferate, and dictates the coupling of $\hat{f}$-fermions to $\hat{\tau}$. To understand this, we perform another Berry-phase analysis. As before, our starting point is an open domain wall end, see Fig.~\ref{fig:fig4}~(a). Across the domain wall, the continuous $\bO$ field experiences a $\pi$-phase slip; in a continuum description, this can be described by an in-plane rotation of the $\bO$ field on scales comparable to the lattice constant, illustrated by grey arrows in Fig.~\ref{fig:fig4}~(a). Therefore, when a magnetic polaron encircles the end of an open domain wall string its spin orientation adiabatically follows the direction of the $\bO$ field, and correspondingly it picks up a Berry phase of $\varphi_B = \pi$ [Fig.~\ref{fig:fig4}~(a)]. Meanwhile, closed paths that do not encircle the end of an open string result in a vanishing Berry phase. 

Thus we find that magnetic polarons and open domain wall ends are mutual semions. Since we identified the latter with $\zz_2$ magnetic excitations (visons) of the fluctuating string field $\hat{\tau}$, we conclude that the fermions $\hat{f}$ describing magnetic polarons must carry the corresponding $\zz_2$ electric charge: The mutual semionic braiding statistics of $\zz_2$ electric and magnetic excitations reflects the topological $\pi$-Berry phase of magnetic polarons that needs to be taken into account when expressing the microscopic field $\bO$ by the AFM domain-wall strings $\hat{\tau}$.

These new insights lead us to the following effective Hamiltonian capturing the motion of magnetic polarons in a quantum loop gas of fluctuating domain walls,
    \begin{equation}
\begin{aligned}
    \hat{\mathcal{H}} =  \hat{\mathcal{H}}_{\text{TC}} +  \sum_{\substack{\bbrakket{\i, \k}\\\j\in \text{NN}\{\i,\k\}}} \Big[ t^{\text{MP}}_{\bbrakket{\i,\k}} \hat{f}^{\dagger}_{\i, \tilde{\sigma}} \hat{\tau}^z_{\bell = \braket{\i,\j}}  \hat{\tau}^z_{\bell = \braket{\j,\k}} \hat{f}_{\k, \tilde{\sigma}}^{\phantom\dagger} +\text{H.c.} \Big].
\end{aligned}
\label{eq:LGT}
\end{equation}
Here, fermionic operators $\hat{f}_{\i}^{(\dagger)}$ create (remove) magnetic polarons on site $\i$ and $\bbrakket{\i,\k}$ are NNN pairs on the square lattice; the sum over $\j$ includes both NNN paths when coupling sites $\i \leftrightarrow \k$ with hopping strength $t^{\text{MP}}_{\bbrakket{\i,\k}}$. The latter is proportional to the superexchange energy of the bare Hubbard model, $t^2/U$~\cite{Grusdt2018_partons}. 
We note that there is a fundamental reason why fermions $\hat{f}$ with different sublattice indices do not mix: Hopping between sites with different sublattice indices is not invariant under the combined action of translation and time-reversal operations and is therefore prohibited—as we show explicitly in Appendix~\ref{sec:appA}.

To capture the semionic statistics between magnetic polarons and open string ends, the polarons couple to electric field lines $\hat{\tau}^z$ when hopping on the lattice. In particular, this coupling to gauge fluctuations does not disrupt the fermionic quasiparticle nature of magnetic polarons. Consequently, both fermionic quasiparticles and the  topological quantum field theory (TQFT) describing the $\zz_2$ spin liquid contribute with a direct sum to Oshikawa's momentum balance argument~\cite{Oshikawa2000}. Repeating the arguments for an FL* from Refs.~\cite{Senthil2003, sachdev2016}, the spin liquid absorbs flux that corresponds to unit density, which results in the formation of a small Fermi surface with carrier density $\delta$ constituted by magnetic polarons.

The above arguments are akin to the doped quantum dimer scenario~\cite{Chowdhury2014, Punk2015, sachdev2016}, where the background spins form a quantum spin liquid that preserves the spin's underlying SU(2) symmetry. In this scenario, holon-spinon bound states that carry the same quantum numbers as fermionic holes then form a small Fermi surface. We emphasize that in our case, the background spins break the SU(2) symmetry. Their order is not disrupted by magnetic frustration but instead by fluctuating domain walls, which form a string-net condensate with $\zz_2$ topological order and fractionalized excitations: The broken symmetry together with the sublattice fluctuations defines the geometric nature of our scenario.

Moreover, in our scenario, the fermions $\hat{f}$ carry a $\mathbb{Z}_2$ gauge charge (in contrast to, for example, the gauge-neutral fermions in doped quantum dimer models), thereby rendering our model an orthogonal metal~\cite{Ruegg2010, Nandkishore2012}. The connection to OMs in Eq.~\eqref{eq:LGT} becomes particularly clear upon introducing a bosonic Higgs-like field on lattice sites $\mathbf{i}$, which also carries a $\mathbb{Z}_2$ gauge charge; the physical (gauge-invariant) electron is then a composite of this field and the gauge-charged fermions. As shown in Appendix~\ref{sec:appB}, this leads to a Hamiltonian structure closely analogous to that explored in Ref.~\cite{Gazit2020}. Due to the orthogonality of the gauge-charged fermions $\hat{f}$ to the physical electrons, the ground state---devoid of Higgs excitations---exhibits a vanishing photoemission signal. However, at elevated temperatures---relevant for current experiments---a finite density of thermally excited Higgs bosons can result in a nonzero, albeit incoherent, signal in ARPES measurements. In particular, discernible signals are expected when the temperature exceeds the energy scale of the Higgs gap (which vanishes at criticality where the toric code undergoes a transition from a topologically ordered to a trivial phase), see Appendix~\ref{sec:appB}. This is underlined by quantum Monte Carlo simulations in Ref.~\cite{Gazit2020}, where broad, arc-like features in the electron spectral function appear in an effective model of orthogonal metals. Importantly, other observables such as transport and quantum oscillation measurements remain insensitive to the orthogonality of the fermions $\hat{f}$, as they probe the charge and momentum carried by these quasiparticles, not their overlap with the physical electron operator. 

\subsection{Hidden quantum critical point (hQCP)} When suppressing superconductivity in hole-doped cuprates using strong magnetic fields, the pseudogap persists at low temperatures up to a critical doping of $\delta_c \sim 19 \%$~\cite{Sachdev2010, Badoux2016}. Around this doping level, non-Fermi liquid transport characteristics have been observed, whereby the resistivity grows linearly with temperature (strange metal)~\cite{Hussey2008}. The location of the strange metal phase in the phase diagram suggests that the non-Fermi liquid behavior is caused by a quantum critical point~\cite{Sachdev1992}, where the pseudogap metal becomes unstable and turns into a plain-vanilla Fermi liquid, see Fig.~\ref{fig:fig1}. This transition is accompanied by a change of the Fermi surface volume, and is hence believed to be of topological nature. 

Within our fluctuating domain wall picture, we propose the following scenario for the appearance of a quantum critical point as a function of doping: For low dopings, in the pseudogap regime, the ground state is characterized by hidden AFM order in (the 2D generalization of) squeezed space, i.e. the state spontaneously breaks the SU(2) symmetry but long-range AFM correlations are hidden by the proliferation of AFM domain walls. Eventually the SU(2) symmetry of the underlying spins must be fully restored, as in the Fermi liquid (with large Fermi surface) observed beyond $\delta_c \sim 19 \%$ doping. In between, this necessitates the existence of a quantum critical point associated with the formation of the SU(2) broken AFM in squeezed space. Due to the nature of correlations in the original lattice model, which are short-ranged on both sides of this transition, the quantum critical point itself is ``hidden'' (hQCP) and cannot be directly associated with a diverging correlation length. 

To understand the microscopic origin of the hQCP, let us consider the effective Hamiltonian describing spins in 2D squeezed space. Since we work in the strong- coupling regime of the Hubbard model, the tunneling energy of the mobile dopants, $t$, exceeds the AFM super-exchange coupling, $J$, which introduces significant charge fluctuations. Both, for dopants bound into stripe-like structures as well as mobile holes forming magnetic polarons, this has been argued to create significant frustration in the surrounding spin background~\cite{Anderson1987, Zaanen2001, Grusdt2018_partons, Grusdt2019_string}. In a quantitative analysis in a mixed-dimensional setting, the effective spin system in squeezed space has been demonstrated to be accurately described by a $J_1$-$J_2$-type Heisenberg antiferromagnet using Hamiltonian reconstruction schemes, with a ratio $J_2/J_1$ that is doping dependent~\cite{Schloemer2023_quantifying}. It was argued that as hole-doping increases, the effective squeezed spin system can be driven into a highly frustrated phase in the $J_1$-$J_2$ phase diagram, whose nature has been under active investigation and remains a promising candidate for the realization of a quantum spin liquid~\cite{Anderson1973, White1994, Mezzacapo2012, Hu2013, Wang2013, Gong2014, Jiang2014, Haghshendas2018}. 

We propose that at the critical doping value $\delta_c\sim 19\%$, the frustrating effect of charge fluctuations becomes too large. In terms of our field theory, this means that the spin stiffness $\rho_s$ associated with the NLSM describing the AFM $\tilde{\bO}$ in squeezed space, Eq.~\eqref{eqZcl}, becomes too small: quantum fluctuations restore the hidden SU(2) symmetry. Without the diverging correlation length in squeezed space, the fluctuating domain wall field $\tau^z$, Eq.~\eqref{eq:tauz}, becomes ill-defined and our field theoretic description is no longer valid at higher dopings.

The hQCP itself can only be detected through highly non-local observables that capture correlations in squeezed space. However, non-Fermi liquid behavior around the critical point can still be observed in transport measurements, consistent with extensive experimental evidence~\cite{Michon2021, Arpaia2023}. In particular, we expect signatures of a quantum critical fan emerging from the hQCP with increasing temperature. It is also conceivable that right at the hQCP the spins in squeezed space form an SU(2) symmetric quantum spin liquid, turning magnetic polarons into a more conventional (non-geometric) FL* before topological order is lost in the FL regime. This scenario is reminiscent of the physics of a doped quantum dimer model featuring an FL* at the Rokhsar-Kivelson point~\cite{Punk2015}, and suggests that the hQCP itself may be described by the field theory of an FL*~\cite{Senthil2003}.

This points to an intricate relationship between QCP signatures in hole-doped cuprates and those found in electron-doped cuprates~\cite{Armitage2010}, heavy-fermion metals~\cite{Gegenwart2008}, iron-based superconductors~\cite{Stewart2011}, and quasi-1D organic conductors~\cite{Jerome2024}. In the latter systems, the QCP scenario is well established, where the AFM phase terminates and a $d$-wave superconducting dome emerges around the QCP~\cite{Paschen2004, Monthoux2007, Taillefer2010, Proust2010}. In hole-doped cuprates, while strong evidence suggests that the pseudogap critical point is a QCP based on specific heat and transport measurements, no diverging length scale related to quantum criticality has been identified thus far~\cite{Proust2010}. In our domain wall scenario, hidden AFM correlations are implicated in the formation of the pseudogap in hole-doped cuprates. We propose that these fluctuations lead to the hQCP located within the $T_c$ dome, which potentially offers new insights into the origin of superconductivity in these materials, driven by AFM spin fluctuations in squeezed space. This includes scenarios where superconductivity has been proposed to arise from magnon exchange~\cite{Miyake1986, Scalapino1986}, spin-bag mechanism~\cite{Su1988, Schrieffer1988} or an emergent Feshbach resonance~\cite{Homeier2023}.

\section{Discussion}
We propose a theoretical framework for hole-doped cuprates where fluctuations of closed, stripe-like structures give rise to hidden AFM order. At elevated temperatures and very low doping $\delta\sim 5\%$, we suggest that a transition occurs from long-range AFM to hidden order, driven by an Ising* criticality where string nets percolate. In the ground state, we argue that quantum fluctuations extend the hidden order phase down to $T=0$, with topological properties naturally emerging from a minimal model -- leading to magnetic (or spin-) polarons forming a small Fermi surface coupled to topological excitations of a string-net condensate. At a critical doping, we propose that hidden order vanishes, resulting in a transition from a geometric orthogonal metal (GOM) to a conventional Fermi liquid at a hidden quantum critical point (hQCP), which furthermore constitutes a scenario that explains signatures of quantum criticality observed in hole-doped cuprates. These results are summarized in Fig.~\ref{fig:fig1}.

Our work unifies several puzzles in the cuprates, including the relation of the pseudogap phase to stripes and antiferromagnetism, and paves the way for intriguing future directions. One avenue involves exploring half-vortex excitations linked by domain wall strings (open string ends), which could drive a Berezinskii-Kosterlitz-Thouless (BKT)-type transition; we speculate that this might be related to the physics associated with the temperature scale $T^*$ revealed in various observables~\cite{Norman2005, Keimer2015, Timusk1999, Lee2006, Chowdhury2015}. Running semi-classical numerical simulations that incorporate their structure and interactions may shed light on the fate of hidden order at higher temperatures. Another promising direction is to quantify the frustration in squeezed space caused by fluctuations of closed loops within the hidden order phase. By employing Hamiltonian reconstruction methods~\cite{Schloemer2023_quantifying}, this could provide deeper insights into the nature of a possible quantum phase transition at the hidden critical doping $\delta_c$.

In the hidden order phase, the SU(2) symmetry of the underlying local moments is spontaneously broken, possibly explaining signatures of time-reversal symmetry-breaking in certain families of cuprates~\cite{Fauque2006, Li2008Unusual, Xia2008}. This scenario is supported by the observation of pronounced paramagnon peaks up to high doping values beyond the pseudogap regime~\cite{LeTacon2011, Dean2013, Guarise2014}. Our scenario captures key aspects of anomalous transport and quantum oscillation experiments in the pseudogap, and is also generally consistent with ARPES measurements performed at finite temperatures where spectral features resembling Fermi arcs can be expected in OMs, see Appendix~B and Ref.~\cite{Gazit2020}. We hence propose that the origin of Fermi-arcs and the suppression of spectral weight on their backsides has a more subtle origin, which future experiments with high energy and momentum resolution may help to resolve~\cite{Kurokawa2023}.

Our picture of fluctuating stripe-like structures differs fundamentally from other proposals in e.g. Ref.~\cite{Laliberte2011}, where thermodynamic and transport anomalies in cuprates are attributed to Fermi surface reconstruction by static stripe order (see also Ref.~\cite{Vojta2012}). Since quantum oscillations and d.c. transport measurements are essentially static probes, fluctuating stripes (in terms of symmetry-breaking order) likely do not lead to observable features in the pseudogap phase of cuprates. In contrast, in our scenario, these features naturally arise from the topological geometric structure of fluctuating stripes.

Ultracold atom simulation platforms offer a unique opportunity to test our hidden order scenario. By analyzing spin- and charge-resolved snapshots, evidence of our extended definition of squeezed space in two dimensions could be explored. With recent advances in state preparation schemes, achieving cryogenic temperatures of $T/t \lesssim 0.1$ is now within reach for analog quantum simulators~\cite{Xu2025}. This advancement will enable the study of regimes where stripes melt in 2D settings, allowing our hypothesis of fluctuating domain walls causing the peculiar pseudogap physics to be rigorously tested. If such patterns are observed, the real-space nature of many-body snapshots allows for a direct calculation of spin-spin correlations within squeezed space~\cite{Hilker2017}, which could be used to identify the transition from the pseudogap to the Fermi liquid phase~\cite{Koepsell2021}. Moreover, digital mirror devices (DMDs) can be used to effectively decouple a 2D system into two parts. Sharp boundaries are expected to expel domain wall lines, leading to the reappearance of long-range spin-spin correlations. True long-range order is anticipated to emerge along these 1D boundary lines, contrasting the power-law decay expected in 1D systems. In this context, exploring potential links between the pseudogap phase and symmetry-protected topological phases is interesting~\cite{Wilke2025}. We note that this predicted re-emergence of long-range spin order at sharp boundaries is, in principle, also accessible in solid state systems through detailed investigations of magnetic order and excitations at these interfaces. This will constitute the ultimate test of our proposed theoretical scenario. \\
     
\textbf{Acknowledgments.---} We thank Pit Bermes, Immanuel Bloch, Pietro M. Bonetti, Eugene Demler, Gesa D\"unnweber, Antoine Georges, Lukas Homeier, Hannah Lange, Simon Linsel, Philip Phillips, Lode Pollet, Subir Sachdev and Alexander Wietek for fruitful discussions. This research was funded by the Deutsche Forschungsgemeinschaft (DFG, German Research Foundation) under Germany's Excellence Strategy -- EXC-2111 -- 390814868 and has received funding from the European Research Council (ERC) under the European Union’s Horizon 2020 research and innovation programm (Grant Agreement no 948141) — ERC Starting Grant SimUcQuam. 

\appendix
\section{Sub-lattice mixing}
\label{sec:appA}
In the main text, we motivated our effective low-energy theory of the FH model, given by Eq.~\eqref{eq:LGT}. In this framework, we have argued that magnetic polarons hop between NNN sites and acquire a $\zz_2$ charge through their coupling to the gauge field $\hat{\tau}^z$ describing a string-net of fluctuating stripes (AFM domain walls). 

We now demonstrate that sublattice mixing of magnetic polarons in an AFM background (e.g. given by NN couplings) is indeed prohibited by the combined translational and time-reversal symmetry. To this end, consider first the translation operator $\hat{T}$. To clarify the following arguments, we explicitly include the sublattice index $\alpha$ in the site label, denoting a site as $\i_{\alpha}$. The fermionic operators $\hat{f}_{\i_{\alpha}, \tilde{\sigma}}$ transform under translation by one lattice site, $\i \rightarrow \i + \mathbf{a}$, as follows,
\begin{equation}
    \hat{T} \hat{f}_{\i_{\alpha}, \tilde{\sigma}}\hat{T}^{-1} = \hat{f}_{(\i+\mathbf{a})_{-\alpha}, -\tilde{\sigma}},
\end{equation}
i.e., both the vacuum label $\tilde{\sigma}$ and the sublattice index $\alpha$ switch signs.

Similarly, the time-reversal operator $\hat{\Theta}$, acting on microscopic fermions $\hat{c}_{\i, \sigma}$ as $\hat{\Theta} \hat{c}_{\i, \sigma} \hat{\Theta}^{-1} = \sigma \hat{c}_{\i, -\sigma}$, acts on the $\hat{f}_{\i_{\alpha}, \tilde{\sigma}}$ fermions as
\begin{equation}
    \hat{\Theta} \hat{f}_{\i_{\alpha}, \tilde{\sigma}} \hat{\Theta}^{-1} = \sigma(\alpha, \tilde{\sigma},\Sigma) \hat{f}_{\i_{\alpha}, -\tilde{\sigma}},
\end{equation}
where we have used that $\tilde{\sigma}(\alpha, -\sigma, \Sigma) = -\tilde{\sigma}(\alpha, \sigma, \Sigma)$ by inverting Eq.~\eqref{eq:sigsigp}.

Now, consider a term that couples two different sublattices (for clarity, we omit the site index $\i$), given by $\hat{f}^{\dagger}_{\alpha, \tilde{\sigma}} \hat{f}_{-\alpha, \tilde{\sigma}}^{\vphantom\dagger}$. This term transforms under the combined action of translation $\hat{T}$ and time-reversal $\hat{\Theta}$ as
\begin{equation}
\begin{aligned}
    \hat{T} \hat{\Theta} \hat{f}^{\dagger}_{\alpha, \tilde{\sigma}} \hat{f}_{-\alpha, \tilde{\sigma}}^{\vphantom\dagger} \hat{\Theta}^{-1} \hat{T}^{-1} &= \sigma(\alpha, \tilde{\sigma}, \Sigma) \sigma(-\alpha, \tilde{\sigma}, \Sigma) \hat{f}^{\dagger}_{-\alpha, \tilde{\sigma}} \hat{f}_{\alpha, \tilde{\sigma}}^{\vphantom\dagger} \\ 
    &= - \hat{f}^{\dagger}_{-\alpha, \tilde{\sigma}} \hat{f}_{\alpha, \tilde{\sigma}}^{\vphantom\dagger}.
\end{aligned}
\end{equation}
This result shows that hopping between different sublattices is not invariant under the combined action of translation and time-reversal symmetry. Consequently, in our low-energy description of magnetic polarons coupled to a string net condensate, only couplings within the same sublattice are allowed.

\section{Higgs field and ARPES signal}
\label{sec:appB}

To establish the relation to microscopic models of orthogonal metals as studied e.g. in Ref.~\cite{Gazit2020}, we introduce a Higgs field $\hat{\sigma}$ into the Hamiltonian Eq.~\eqref{eq:LGT}. We focus here on the toric code in the closed-loop subspace with loop tension [i.e., $\lambda = 0$ in Eq.~\eqref{eq:LGT}]. The Higgs field is constructed such that
\begin{equation}
    \hat{\sigma}^x_{\j} \prod_{\bell \in +_\j} \hat{\tau}^x_{\bell} = 1,
\end{equation}
leading to the relations
\begin{equation}
\begin{gathered}
    \prod_{\bell \in +_\j} \hat{\tau}^x_{\bell} = \hat{\sigma}^x_{\j}, \\
    \hat{\tau}^z_{\bell = \braket{\i,\j}} \to \hat{\sigma}^z_{\i} \hat{\tau}^z_{\bell = \braket{\i,\j}} \hat{\sigma}^z_{\j}.
\end{gathered}
\label{eq:higgstrafo}
\end{equation}
Inserting Eq.~\eqref{eq:higgstrafo} into Eq.~\eqref{eq:LGT} yields
\begin{widetext}
    \begin{equation}
\begin{aligned}
    \hat{\mathcal{H}} =  - K_{\square} \sum_{\square}\prod_{\bell \in \square} \hat{\tau}^z_{\bell} - h \sum_{\bell=\braket{\i,\j}} \hat{\sigma}^z_\i \hat{\tau}^z_{\bell=\braket{\i,\j}} \hat{\sigma}^z_\j + K_{+} \sum_{\j}\hat{\sigma}^x_\j +  \sum_{\substack{\bbrakket{\i, \mathbf{m}}\\\j\in \text{NN}\{\i,\mathbf{m}\}}} \Big[ t^{\text{MP}}_{\bbrakket{\i,\k}} \hat{f}^{\dagger}_{\i, \tilde{\sigma}} \hat{\sigma}^z_\i \hat{\tau}^z_{\bell = \braket{\i,\j}}  \hat{\tau}^z_{\bell = \braket{\j,\mathbf{m}}} \hat{\sigma}^z_{\mathbf{m}} \hat{f}_{\mathbf{m}, \tilde{\sigma}}^{\phantom\dagger} +\text{H.c.} \Big],
\end{aligned}
\label{eq:LGT_higgs}
\end{equation}
\end{widetext}
where the operator $\hat{f}^{\dagger}_{\i, \tilde{\sigma}} \hat{\sigma}^z_\i$ creates a gauge-invariant fermion.

Quasiparticle excitations can now be approximated by gauge-invariant fermion-Higgs composites hopping on sites $\i_{\tilde{\sigma}}$ on one sublattice,
\begin{equation}
    \hat{\gamma}_{\k,\tilde{\sigma}} = \frac{1}{\sqrt{N/2}} \sum_{\i_{\tilde{\sigma}}} e^{i \k \cdot \i_{\tilde{\sigma}}} \left( \prod_{\bell \in \mathcal{L}_{\i_{\tilde{\sigma}}}} \hat{\tau}^z_{\bell} \right) \hat{\sigma}^z_{\i_{\tilde{\sigma}}} \hat{f}_{\i_{\tilde{\sigma}},\tilde{\sigma}}.
\end{equation}
Here, $\k$ is the momentum in the magnetic BZ, and $\mathcal{L}_{\i}$ denotes a path from a reference site at infinity to site $\i$; due to the closed-loop constraint, the product of $\hat{\tau}^z$ operators along $\mathcal{L}_{\i}$ is independent of the precise path taken.  

For the quasiparticle weight $Z$ to be finite, a finite density of gapped Higgs excitations must therefore be present. This is illustrated by considering a local state with a single Higgs excitation,
\begin{equation}
    \ket{\Phi_\i} = \left( \prod_{\bell \in \mathcal{L}_{\i}} \hat{\tau}^z_{\bell} \right) \hat{\sigma}^z_\i \ket{\Psi_0},
\end{equation}
where $\ket{\Psi_0}$ is the ground state of the (undoped) perturbed toric code.

As a result, the broadening of the fermionic spectral features in momentum space, $\Delta k$, is related to the correlation length $\xi$ of the Higgs field correlator,
\begin{equation}
    C^{zz}(\mathbf{d}) = \left\langle \Psi_0 \Big| \hat{\sigma}^z_\j 
    \left( \prod_{\bell \in \mathcal{L}_{\j, \j+\mathbf{d}}} \hat{\tau}^z_{\bell} \right) 
    \hat{\sigma}^z_{\j+\mathbf{d}} \Big| \Psi_0 \right\rangle \simeq e^{- d/\xi},
\end{equation}
i.e. $\Delta k \simeq 1/\xi$. In the deconfined (topologically non-trivial) phase, the finite Higgs gap leads to exponentially decaying correlations, and thus the quasi-particle weight is heavily suppressed. 

In contrast, in the topologically trivial (Higgs) phase, $C^{zz}(\mathbf{d})$ becomes long-ranged; at criticality, Higgs excitations become gapless. Near criticality but in the topologically non-trivial phase, though decaying exponentially, the correlations become significantly extended. As a result, it is possible at finite temperatures---when the thermal energy exceeds the Higgs gap---to obtain ARPES features which are broadened but not entirely suppressed, in contrast to the complete absence of spectral weight expected in the $T=0$ ground state in the deconfined phase. This picture aligns with quantum Monte Carlo simulations in Ref.~\cite{Gazit2020}, which show broad, arc-like features in the electron spectral function of orthogonal metals at finite temperatures.

\section{Effective action}
We here explicitly formulate the effective action of our theory consisting of an odd $\zz_2$ quantum spin liquid (string-net condensate) coupled to magnetic polarons $\hat{f}$, Eq.~\eqref{eq:LGT}. Following the standard Fradkin--Shenker procedure (see, e.g., Refs.~\cite{FradkinShenker1979, Sachdev2019_review}), we extend our original \(\mathbb{Z}_2\) gauge theory to a U(1) gauge theory. In this construction, adding terms that explicitly break the symmetry back down to \(\mathbb{Z}_2\) is not allowed, since they would violate U(1) gauge invariance. Instead, a dynamical Higgs field $H$ is introduced on the lattice sites that transforms as a charge-2 scalar field under U(1).

By appropriately modifying the Gauss law constraint and enforcing it via Lagrange multipliers on the temporal component \(A_\tau\) of the U(1) gauge field, the theory becomes a relativistic model of the Higgs field and fermionic matter coupled to the U(1) gauge field \(A_\mu\) (with \(\mu=x,y,z\)). In particular, the effective action is given by
\begin{widetext}
\begin{equation}
\begin{aligned}
    &\mathcal{S} = \left(\int d\mathbf{x}\, \mathcal{L}_H\right) + \mathcal{S}_B + \mathcal{S}_{\text{monopole}} + \mathcal{S}_f,\\[1mm]
    &\mathcal{L}_H = \left|(\partial_\mu - 2iA_\mu) H \right|^2 + g\, |H|^2 + u\, |H|^4 + K_{\square} \left(\epsilon_{\mu \nu \lambda} \partial_\nu A_\lambda \right)^2,\\[1mm]
    &\mathcal{S}_B = i\sum_{\i} \eta_\i \int d\tau\, A_{\i\tau},\\[1mm]
    &\mathcal{S}_{\text{monopole}} = \sum_\i \int d\tau\, \mathcal{L}_{\text{monopole}},\\[1mm]
    &\mathcal{S}_f = \beta \sum_{\substack{\i, \alpha, \alpha' \\ \i +\mathbf{e}_{\alpha} + \mathbf{e}_{\alpha'}\neq \i}} \Big[ t^{\text{MP}}_{\langle \i,\i + \mathbf{e}_{\alpha} + \mathbf{e}_{\alpha'}\rangle} f^*_{\i, \tilde{\sigma}}\, e^{i\eta_\i A_{\i, \alpha}}\,  e^{i\eta_{\i + \mathbf{e}_\alpha} A_{\i+\mathbf{e}_{\alpha}, \alpha'}} f_{\i+\mathbf{e}_{\alpha} + \mathbf{e}_{\alpha'}, \tilde{\sigma}} +\text{H.c.} \Big].
\end{aligned}
\label{eq:Seff}
\end{equation}
\end{widetext}
In this expression the Higgs term is taken in the continuum limit while the remaining terms are evaluated on the lattice. In the latter, $A_{\i \alpha}$ corresonds to the gauge field associated with sites $\i, \i+\mathbf{e}_{\alpha}$, with $\mathbf{e}_{\alpha}$ the unit vectors in $\alpha=x,y$ directions. In particular, the factor $\eta_\i = (-1)^{i_x+i_y}$ is introduced in the definition of the U(1) gauge field
\begin{equation}
\hat{\tau}^z_{\bell = \braket{\i, \i + \mathbf{e}_{\alpha}}} = e^{i\eta_\i A_{\i \alpha}},
\end{equation}
to recast the original flux term into the form of a lattice curl (which in the continuum limit gives rise to the last term $\propto K_{\square}$ in \(\mathcal{L}_H\))~\cite{Sachdev2019_review}. The Berry phase term \(\mathcal{S}_B\) explicitly enforces the odd gauge constraint, which in the U(1) theory reads
\begin{equation}
\Delta_{\alpha} E_{\i\alpha} - 2\hat{N}_\i = \eta_\i,
\end{equation}
where \(\hat{N}_\i\) is the density operator for the Higgs field and the discrete lattice derivative is defined as
\begin{equation}
\Delta_\alpha E_{\i\alpha} = E_{\i+\mathbf{e}_\alpha} - E_{\i},
\end{equation}
with the canonical commutation relation \([A_{\i \alpha}, E_{\j \beta}] = i\delta_{\i\j}\delta_{\alpha\beta}\). Because the fluctuations of the gauge field are significant, we also include Dirac monopole instantons---configurations in which the U(1) flux changes by \(2\pi\)---schematically represented by \(\mathcal{S}_{\text{monopole}}\)~\cite{Sachdev2019_review}.

Gauge invariance of the effective action is maintained under the transformations
\begin{equation}
A_{i\alpha} \rightarrow A_{i\alpha} + \Delta_\alpha f_i,\quad H_i \rightarrow H_i\, e^{2if_i},
\end{equation}
or, in the continuum,
\begin{equation}
A_\mu \rightarrow A_\mu + \partial_\mu f,\quad H \rightarrow H\, e^{2if}.
\end{equation}

We finally note that, as the fermionic degrees of freedom couple to the gauge field, the corresponding Berry phases of Fig.~\ref{fig:fig4} are incorporated in the theory Eq.~\eqref{eq:Seff} by construction.

\bibliography{bibliography_GeoFL.bib}

\end{document}